\documentclass[floats,preprint,prl,aps]{revtex4}
\usepackage{graphicx}

\newcommand\beq{\begin{equation}}
\newcommand\eeq{\end{equation}}
\newcommand\bea{\begin{eqnarray}}
\newcommand\eea{\end{eqnarray}}
\newcommand\non{\nonumber}
\newcommand\bib{\bibitem}

\begin{document}


\title{\bf Quantum phase diagram of a spin-1/2 antiferromagnetic chain
with magnetic impurity}

\author{\bf Sujit Sarkar}
\address{\it PoornaPrajna Institute of Scientific Research,
4 Sadashivanagar, Bangalore 5600 80, India.\\
E-Mail: sujit@physics.iisc.ernet.in\\
Phone: 091 80 23612511/23619034, Fax: 091-80-2360-0228 \\}
\date{\today}

\begin{abstract}
We present the renormalization group (RG) flow diagram of a spin-half
antiferromagnetic chain with magnetic impurity and one altered link. In
this two parameters (competing interactions) model, one can find
the complex phase diagram with many interesting fixed points.
There is no evidence of intermediate stable fixed point in weak coupling phase.
It may arise at the strong coupling phase.
Depending on the strength of couplings the phases correspond
either to a decoupled spin with Curie law behavior or a logarithmically
diverging impurity susceptibility as in the two channel Kondo problem.
Pacs: 75.10.Pq,
05.10.Cc
  
\end{abstract}
\maketitle


\section{I. Introduction}
The physical behaviors of impurities in the low dimensional magnetic and electronic 
systems are interesting in their own right. There are few important studies
in the literature to describe the behavior of different magnetic impurity 
configurations and defects in the 
antiferromagnetic Heisenberg spin chain \cite{aff1,egg1,egg2,egg3}.
The chain with one altered link and two altered links are, respectively 
renormalized to an effective open boundary and periodic 
boundary conditions \cite{egg1}.
We would like to revisit the problem of 
spin-1/2 antiferromagnetic chain with magnetic impurity and 
one altered link (Fig. 1). 
Motivation of this work comes from the considerable amount of debate on the
RG flow diagram and the nature of the Fixed points of this 
problem \cite{egg3,zvy,egg4} and also for the 
interesting physics of low dimensional spin systems \cite{dago}. 
In this communication, 
we would like to
resolve this debate through the numerical analysis of the RG equations
and the stability analysis of the fixed point (FP).  
Before we proceed
further, we would like to state the important results that have already 
existed in the literature of this field.  
An impurity
spin s coupled to one site in the chain gets screened with a decoupled singlet
of spin s- 1/2 and becomes an open chain with one site removed \cite{nag}. 
An impurity 
spin coupled to two sites in the chain is equivalent to the two channel
Kondo effect \cite{gia1}.
Kondo effect in one dimension has developed based on the separation of
charge and spin in the one dimensional electron gas. The single impurity
Kondo effect only involves the spin degree of freedom of the one dimensional
electron gas, the charge degrees of freedom are not playing the fundamental
role in the Kondo effect \cite{nag}. The spin degrees of freedom of the one-dimensional
interacting electron system at low energies can be described as half-integer
Heisenberg antiferromagnetic spin chain. Hence it is natural to look for a
Kondo effect involving a magnetic impurity interacting with Heisenberg chain \cite{egg1}.\\
\section{II. Renormalization Group study of model Hamiltonian} 
We now present renormalization group (RG) study of two parameter 
model (${J_1},{J_2}$). 
The model Hamiltonian of our system is
\beq
H =  J \sum_{i=1}^{N-1} \vec{S_i} . \vec{S_{i+1}} 
+ J_1 \vec{S_0}. (\vec{S_N} + \vec{S_1} ) 
+ J_2  \vec{S_N} . \vec{S_1}  
\label{ham1}
\eeq
$J$ is the nearest-neighbor Heisenberg exchange coupling.
$J_1$, the symmetric coupling of the impurity to two sites
in the chain and the coupling between two sites is $J_2$ (Fig. 1).
At low temperature, this system is known to be well described 
by a level 1 Wess-Zumino-Witten model with a marginal irrelevant
operator \cite{egg1,aff2}. A spin operator at the position x in the chain can be 
expressed in terms of current operators $\vec{J}$ and WZW field $g$,
\beq
\vec{{S_j}} (x) ~=~ \vec{J_L} ~+ ~\vec{J_R} + constant * ~ {(-1)^j} tr [ \vec{\sigma} g]
\eeq 
where $\vec{J_L}$ and $\vec{J_R}$ are the left and right SU(2) currents.
$ {J_L} (x) ~=~ - \frac{i}{4 \sqrt{\pi} } 
tr [ g^{\dagger} {{\partial}_{-}} g {\sigma}] $
$ {J_R} (x) ~=~ \frac{i}{4 \sqrt{\pi} }
tr [{{\partial}_{+}} g g^{\dagger} {\sigma}] $. $g$ is related
with the Abelian boson field $\phi$ and ${\tilde{\phi}}$ \cite{egg1,aff2},
\[ g \propto \left (\begin{array}{cc}
      i e^{i \sqrt{2 \pi}} {\phi} & e^{i \sqrt{2 \pi}} {\tilde{\phi}} \\
    - e^{- i \sqrt{2 \pi}} {\tilde{\phi}}  & -i e^{- i \sqrt{2 \pi}} {\phi}
        \end{array} \right ) \],
where ${\phi}~=~ {\phi}_R  + {\phi}_L $ and $\tilde{{\phi}}~=~ {\phi}_R  - {\phi}_L $.
This model Hamiltonian has already been studied by 
Eggert $et~al.$ \cite{egg3,egg4} by
using 
field-theory arguments and numerical calculations. They have predicted the
possibilities of different fixed points based on simple 
boundary conditions \cite{egg3,egg4}.
Here we briefly describe   
those fixed points and their consequences in the phase diagram. 
(1). $P_{N+1}$: $J_1 = J$ and $J_2 = 0$, 
a periodic chain with $N+1$ sites and no impurity spin. In this fixed
point leading irrelevant operator is $ {{\partial}_x} tr g $ because
the site parity symmetry does not allow more relevant operators. The
authors of Ref. (\cite{egg4} ) conclude that this FP is stable in all directions
of $J_1 - J_2$ phase diagram \cite{egg3,egg4}. We will see in our
study that there is no intermediate stable fixed point.\\
 
(2). $O_N$: $J_1 =0$ and $J_2 = 0$, a periodic chain with N sites and
a decouple impurity spin. The leading operator 
$ (~ \vec{J_L} (0) + \vec{J_L} (N)~). {\vec{S}}_{imp}$ of scaling dimension
one is created by the coupling $J_1$ to the impurity from the open ends.
This operator is marginally relevant for a antiferromagnetic coupling and
irrelevant for ferromagnetic coupling. 
${\vec{S}}_{imp}$ is the spin at the impurity site.
${J_L}, {J_R}$ and $g$ have already defined in Eq. 2.
The coupling between the end spins
, $J_2$, can only produce the irrelevant operators \cite{egg3,egg4}. \\

(3). $O_{N-2}$: ${J_1} = 0$ and ${J_2} \rightarrow \infty$, near this fixed
point, impurity spin is separated by a locked singlet, which is effectively
decoupled from the rest of the chain \cite{egg3,egg4}.\\

(4). $P_N$: $J_1 =0$ and $J_2 =J$, a periodic chain with N sites and a 
decoupled impurity spin. The most relevant operator $tr g$ corresponds to
a slight modification of one link in the chain $J_2$. 
A small coupling to the impurity
spin $J_1$ produces the operator 
$ ( \vec{J_L} + \vec{J_R} ).{\vec{S}}_{imp}$. The 
irrelevant operator ${{\partial}_x} tr ({\vec{\sigma}} g ) $ of dimension
$d =3/2$ is also created by $J_1$ \cite{egg3,egg4}.\\

The author of Ref. \cite{zvy} has argued that the analysis of the FPs of 
Ref. \cite{egg4} 
is not the complete one. There exist several other FPs
like ${J_1} \rightarrow  -\infty$ and ${J_2} \rightarrow -\infty$.
He has argued at least one extra FP exist with 
$J_1 \rightarrow \infty $.\\

The authors of Ref. \cite{egg3,egg4}, have expressed impurity Hamiltonian at the fixed point $P_N$ as 
\beq
H_{imp} = {{\gamma}_1} trg + {{\gamma}_2} ( \vec{J_L} + \vec{J_R} ). \vec{S_{imp}}
+ {{\gamma}_3} {{\partial}_x} tr (\vec{\sigma} g ). \vec{S_{imp}},   
\eeq      
where ${{\gamma}_1} \propto ( J_2 -J)$, ${{\gamma}_2} \propto J_1$
${{\gamma}_3} \propto J_1 $. 
${J_L}, {J_R}$ and $g$ have defined in Eq. 2.
They have obtained the RG equations as
follows 
\bea
\frac{d {{\gamma}_1}}{dl} ~&=&~ \frac{1}{2} {{\gamma}_1} ~-~ \frac{3}{2} 
{{\gamma}_2}{{\gamma}_3},
\non \\
\frac{d {{\gamma}_2}}{dl} ~&=&~ {{\gamma}_2}^{2} - \frac{3}{4} {{\gamma}_3}^{2}  , \non \\
\frac{d {{\gamma}_3}}{dl} ~&=&~ -\frac{1}{2}{{\gamma}_3} + 2 {{\gamma}_3} {{\gamma}_2}   , \non \\
\label{rg1}
\eea
The third equation of the above RG equations has generated dynamically.
We obtain the RG flow diagram by solving the above mentioned equations
with sophisticated numerical package, MATLAB. 
These RG equations have both trivial, 
(${{\gamma}_1}^{*}, {{\gamma}_2}^{*}, {{\gamma}_3}^{*}$) = ($0,0,0$) 
and nontrivial fixed points 
(${{\gamma}_1}^{*}, {{\gamma}_2}^{*}, {{\gamma}_3}^{*}$) = 
($\frac{\sqrt{3}}{8}, \frac{1}{4}, \frac{1}{2 \sqrt{3}} $).
We do the linear stability analysis to check the stability of these fixed points (FP). 
After
the linear stability analysis RG equations reduce to
\beq
\frac{d}{dl} A_1~= ~B_1 A_1 ,
\eeq    
where  
\[ A_1 = \left (\begin{array}{c}
       {\gamma}_1\\{\gamma}_2\\{\gamma}_3
        \end{array} \right ) \] and
\[ B_1 = \left (\begin{array}{ccc}
      \frac{1}{2} & \frac{-3}{2} {{\gamma}_3}^{*}  & \frac{-3}{2} {{\gamma}_2}^{*} \\
       0 &  2  {{\gamma}_2}^{*}  & \frac{-3}{2} {{\gamma}_3}^{*} \\
 0 &  2  {{\gamma}_3}^{*}  & \frac{-1}{2} + 2 {{\gamma}_2}^{*} 
        \end{array} \right ) \].
At the trivial fixed point, $\frac{d {\gamma}_1}{dl}~=~\frac{{\gamma}_1}{2}$ and 
$\frac{d {\gamma}_3}{dl}~=~- \frac{{\gamma}_3}{2}$,
$\frac{d {\gamma}_2}{dl}~=~0 * {\gamma}_2 $.
The equation for ${\gamma}_1$ is unstable whereas the equation for
${\gamma}_3$ is stable. The equation for ${\gamma}_2$ is marginal.
If we look at the next order term for the marginal case, i.e., 
$\frac{d {\gamma}_2}{dl}~=~ a {{\gamma}_2}^2$ ($ a >0 $), we say that FP at 
${\gamma}_2~=~0$
is stable on the $ x<0 $ side and unstable on the $x >0$ side.\\
We now present the stability analysis near to the nontrivial FPs.
After
the linear stability analysis RG equations reduce to
\beq
\frac{d}{dl} A_2~= ~B_2 A_2 ,
\eeq    
where  
\[ A_2 = \left (\begin{array}{c}
       {\gamma}_1\\{\gamma}_2\\{\gamma}_3
        \end{array} \right ) \] and
\[ B_2 = \left (\begin{array}{ccc}
      \frac{1}{2} & \frac{-\sqrt{3}}{4} & \frac{-3}{8} \\
       0 &  \frac{1}{2} & \frac{- \sqrt{3}}{4} \\
 0 &   \frac{1}{\sqrt{3}} & 0 
        \end{array} \right ) \].

The eigenvalues of the matrix $B_2$ are 
(${{\lambda}_1}, {{\lambda}_2}, {{\lambda}_3})~=~(1/2 , 
1/4 + i \sqrt{3} /{4}, 1/4 - i \sqrt{3} /{4}$ ). One of them is
real and positive and the other two have imaginary parts 
, conjugated to each other but
the real part is positive.
Hence the system is in an unstable phase.\\
In Fig. 2, we present the RG flow diagram in ${{\gamma}_1}
-{{\gamma}_2}$ plane, ${{\gamma}_3} =0 $ as a initial parameter.
We observe that here there is no intermediate nontrivial FPs for small
values of coupling constants.
The different coupling constants are flowing off to the higher values
at the different sector of RG flow diagram and the corresponding instabilities
are growing up in the systems. In our RG flow diagram, once the coupling constants
flowing off to the higher values, there is no opportunity to return in the weak
coupling phase. So the system flowing off to the strong coupling phase. 
We conject about the existence of these strong coupling phases following the
prescription of seminal work of Furusaki and Nagaosa 
on one dimensional Kondo problem \cite{nag}, which is
well accepted in the literature. They have had tentatively extended
the scaling equation to the strong coupling region.
One is ${{\gamma}_1} \rightarrow \infty$
(we denote this phase region by A), the effective coupling $J_2$
increases quickly so that two end spins lock into a singlet, upto
a critical value of $J_1$. In region A, system flows to the $O_{N-2}$.
The region B is the another strong coupling phase region, system flows
to the $O_N$. In this phase region effective coupling, ${J_2}$ decreases
to zero. The phase regions, A and B, are upto a critical
value of $J_1$. When $J_1$ exceed the critical value, system drives 
to a another critical region C. This region is another strong coupling
phase region, independent on the initial value of $J_2$. We call this
fixed point as $O_{N+1}$, this FP appears at 
${{\gamma}_1} =0$, i.e., ${J_2} = J$ and 
${J_1} \rightarrow \infty$. 
This fixed point may coincide with the $P_{N+1}$ of Ref. \cite{egg3,egg4}
for large values of $J_{1}$.
There is no evidence of intermediate stable fixed point as claimed in
Ref. [4-5]. There is no stability analysis of FPs in Ref. [4-6], hence
the conjecture regarding the FPs are not consistent ones. 
In this strong coupling regime, 
impurity spin tightly bound to the nearest-neighbor spin and the 
system behaves like a two channel Kondo problem with logarithmic
susceptibility. This FP occurs at $J_2 = J$, so there is no opportunity
that impurity spin and two neighboring spins decoupled from the chain.
This conjecture is consistent with the findings of Ref. \cite{egg4}.
So in this complex phase diagram, there is only one FP,
${O}_{N-2}$, where the spin singlet locked and decouple from the
rest of the chain. In Ref. \cite{egg4}, performed the TMRG calculation
to predict the parabolic phase boundary between the phases $O_N$ and $O_{N-2}$.
The phase boundaries of our study is also parabolic.
Our RG flow diagram is the extensive one
. The phase regions A and B behave as a Curie law behavior
as $T \rightarrow 0$ from the decoupled impurity spin degrees of freedom
and region C is logarithmically divergent impurity susceptibility \cite{gia1}.\\    
In. Fig. 3, we present the RG flow diagram in ${\gamma}_1$ - ${\gamma}_2$
plane, ${{\gamma}_3} = 0.1$ is the initial parameter of the system. We
observe only two strong coupling phases, ${O_{N-2}}$ and ${O_N}$. Most of 
the regions of the phase diagram corresponds to the B phase.\\ 
Similarly in
Fig. 4, ${{\gamma}_3} = -0.1$ is the initial parameter of the system, here
we also observe two strong coupling phase regions like Fig. 3 but 
most of the phase regions covers by the phase A. The studies of the
effect of the initial values of ${\gamma}_3$ are absent in the previous studies
\cite{egg3,zvy,egg4}. There is no evidence of the existence of phase region C.\\  
\section{III. Conclusions}
We have revisited the problem of magnetic impurity in a 
spin-1/2 antiferromagnetic chain with one alter link. 
The phase diagram of the previous studies \cite{egg3,zvy,egg4}
are the schematic ones and there are no sound numerical analyses
to predict the different fixed points and their nature (stability
analysis of the fixed points). 
We have presented  
an extensive RG flow diagram and has also done the stability analysis
of the FPs. We have predicted that there is no evidence of
intermediate stable FPs at the weak coupling limit, i.e.,
for the small values of coupling constants.
FP may arise for the large values of coupling constant (${J_1}$). 
We have concluded that this fixed point does not correspond to any
completely decoupled phase.
We have studied the explicit role of 
${\gamma}_3$ term. 
For 
${{\gamma}_3} = \pm 0.1$, our RG flow diagrams study is entirely new,
the system possessing only two strong coupling phase regions. 
\centerline{\bf Acknowledgments}
\vskip .2 true cm
The author would like to acknowledge The Center for Condensed Matter 
Theory of the 
Physics Department
of IISc for providing working space and also Mr. M. Vasudeva for reading
the manuscript very critically. 

\newpage
\begin{figure}[]
\includegraphics[height=4cm,width=9.0cm,,angle=0]{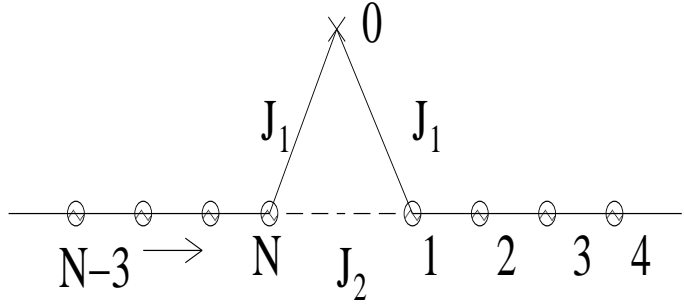}
\caption{Schematic diagram of impurity model with two parameters,
$J_1$ and $J_2$. Cross sign indicates the position of impurity and
the circle represents the regular spin sites in the chain.
}
\label{Fig1}
\end{figure}
\newpage
\begin{figure}[]
\includegraphics[height=7cm,width=9.0cm,,angle=0]{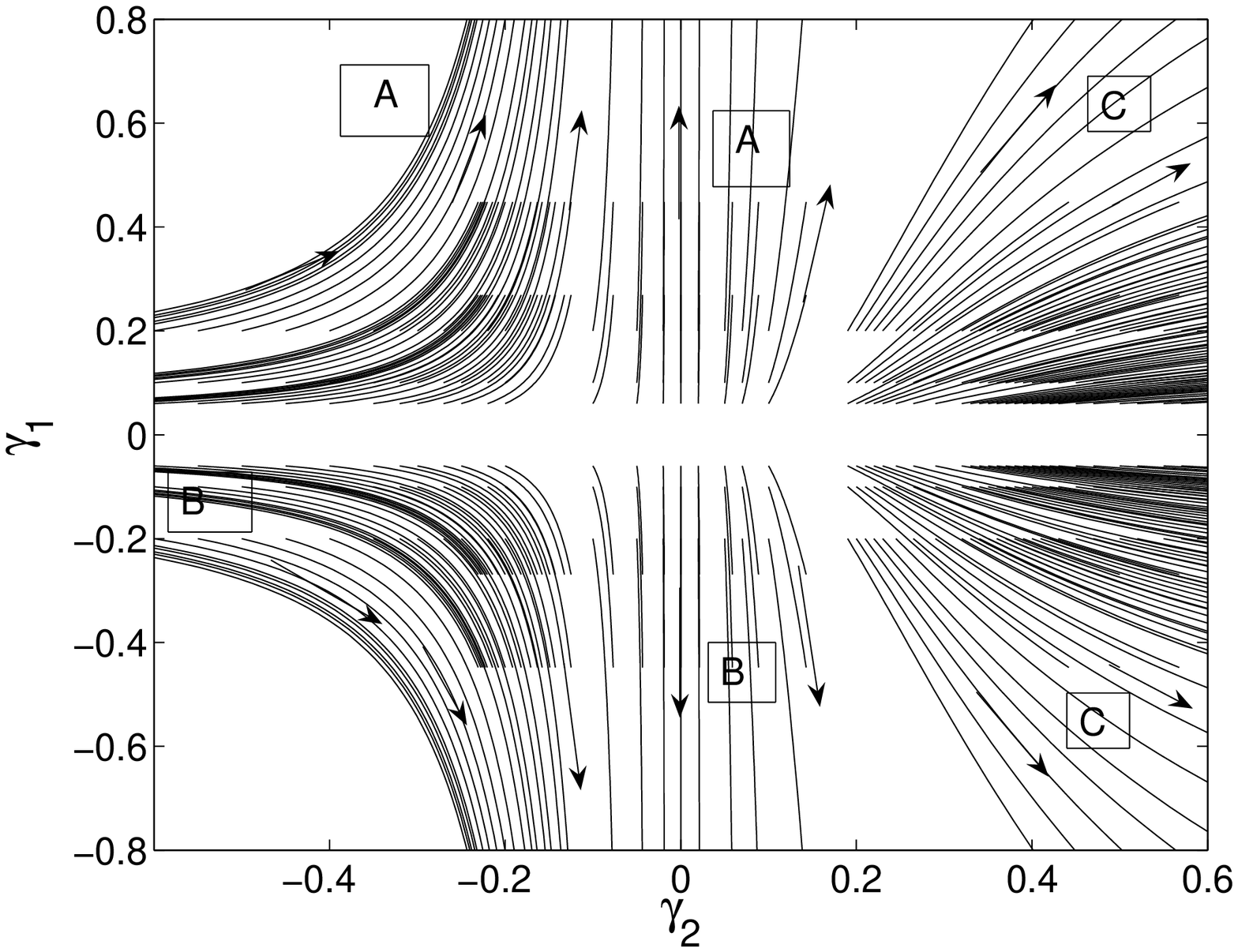}
\caption{The RG flow diagram in the ${\gamma}1 - {\gamma}2$ plane for Eq. 4 . 
The solid line and arrow 
show the flow and the direction respectively. 
This phase diagram consists of three strong coupling phase
regions A ($O_{N-2} $), B ($ O_{N} $) and C ($ O_{N+1} $) (please
see text for detailed analysis).  ${\gamma}3 =0 $.
}
\label{Fig2}
\end{figure}
\newpage
\begin{figure}[]
\includegraphics[height=7cm,width=9.0cm,,angle=0]{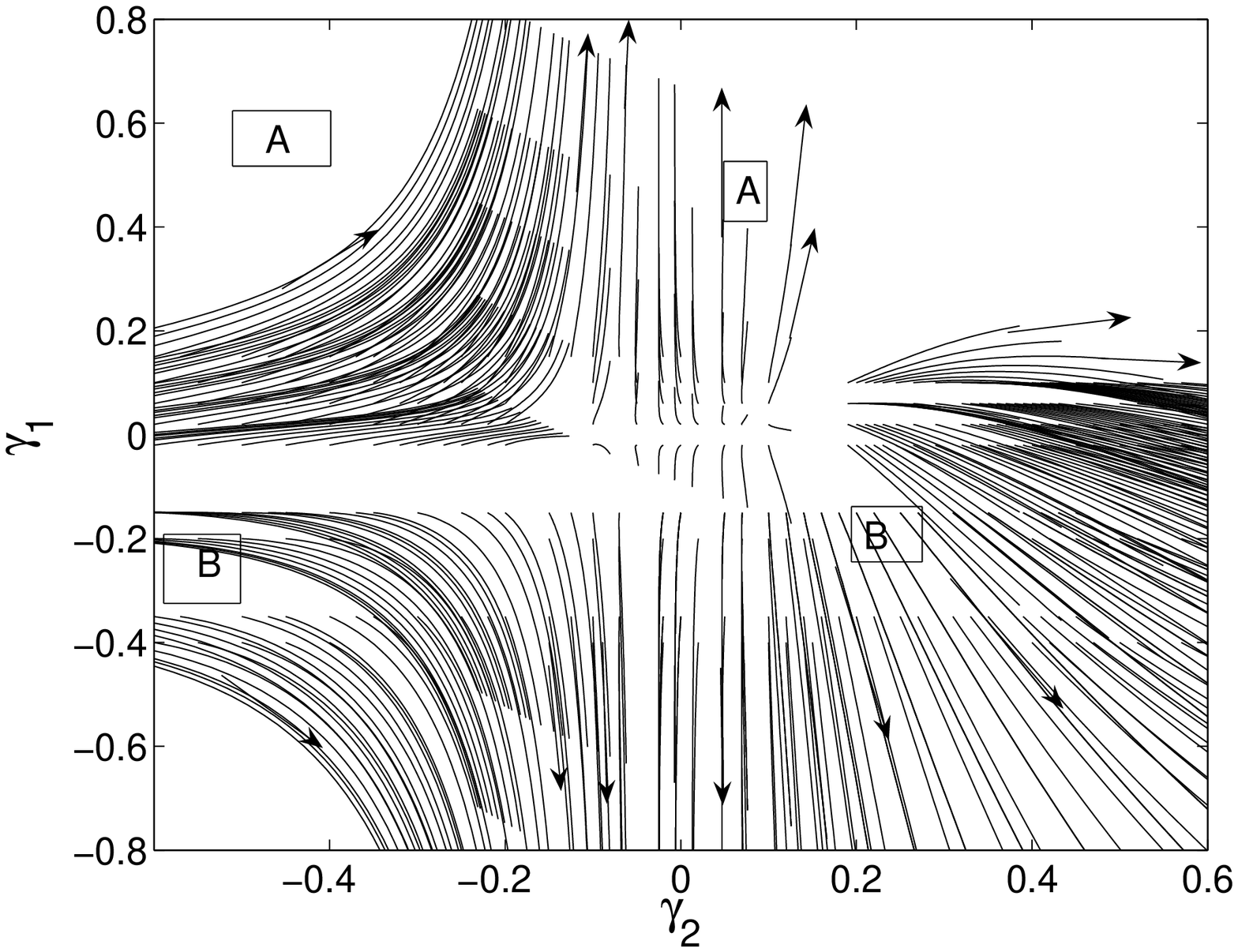}
\caption{The RG flow diagram in the ${\gamma}1 - {\gamma}2$ plane for Eq. 4. 
The solid line and arrow 
show the flow and the direction respectively. 
This phase diagram consists of two strong coupling phase
regions A ($O_{N-2} $) and B ($ O_{N} $) (please
see text for detailed analysis).  ${\gamma}3 =0.1 $.
}
\label{Fig3}
\end{figure}
\newpage
\begin{figure}[]
\includegraphics[height=7cm,width=9.0cm,angle=0]{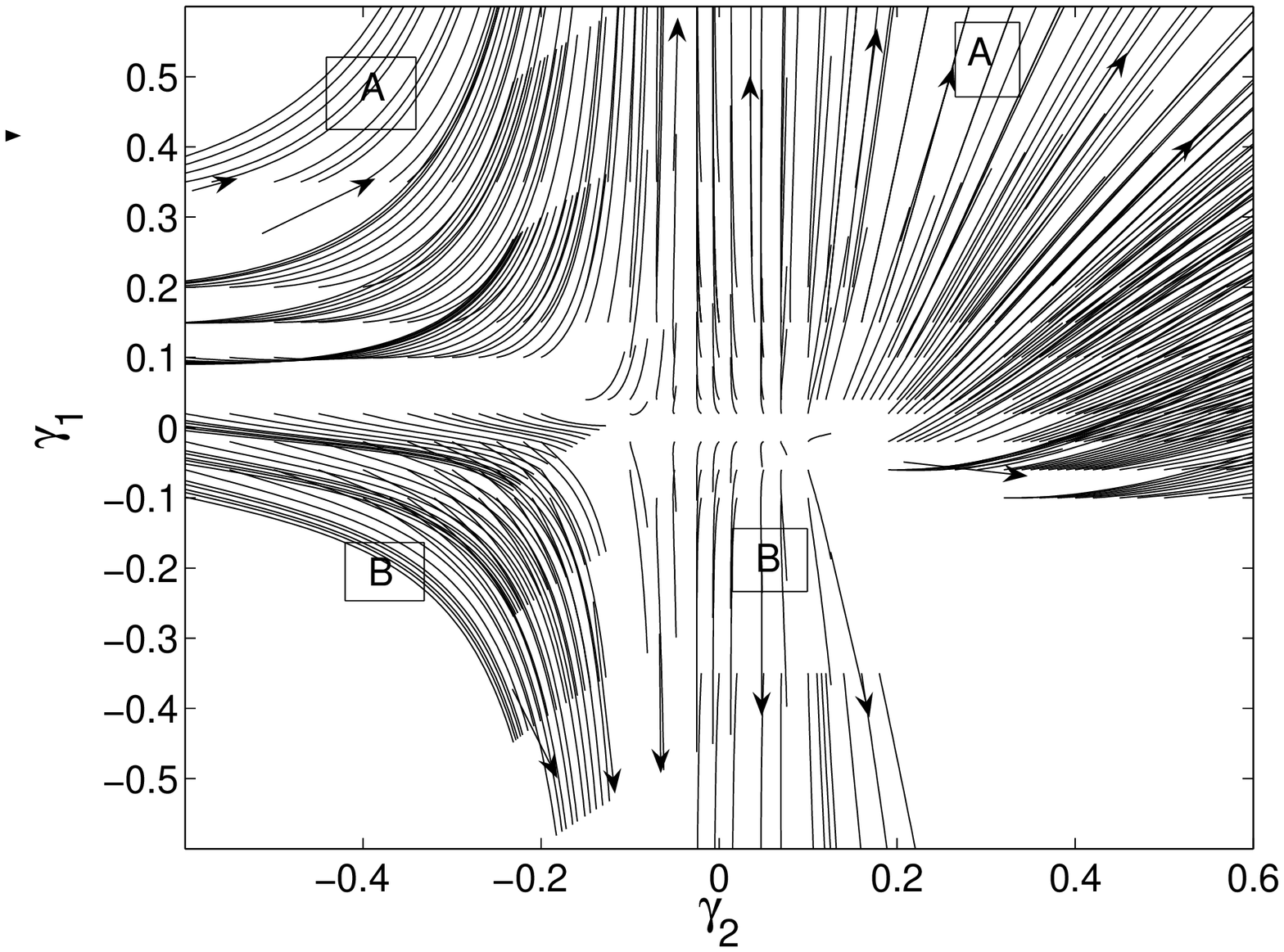}
\caption{The RG flow diagram in the ${\gamma}1 - {\gamma}2$ plane for Eq. 4 . 
The solid line and arrow 
show the flow and the direction respectively. 
This phase diagram consists of two strong coupling phase
regions A ($O_{N-2} $) and B ($ O_{N} $) (please
see text for detailed analysis).  ${\gamma}3 = -0.1 $.
}
\label{Fig4}
\end{figure}
\end{document}